\documentclass[twocolumn,prl,showpacs,preprintnumbers,amsmath,amssymb]{revtex4}

\usepackage{graphicx}
\usepackage{dcolumn}
\usepackage{bm}
 
\begin{document}

\title{New limit on the present temporal variation of the fine structure constant}

\author{E. Peik, B. Lipphardt, H. Schnatz, T. Schneider, Chr. Tamm}
\affiliation{Physikalisch-Technische Bundesanstalt, Bundesallee 100, 38116 Braunschweig, 
Germany}
\author{S. G. Karshenboim}
\affiliation{D. I. Mendeleev Institute for Metrology (VNIIM), 198005 St. Petersburg, Russia\\ and
Max-Planck-Institut f\"ur Quantenoptik, 85748 Garching, Germany}


\begin{abstract}
The comparison of different atomic transition frequencies over time can be used to determine the present value of the temporal derivative of the fine structure constant $\alpha$ in a model-independent way without assumptions on constancy or variability of other parameters, allowing tests of the consequences of unification theories.
We have measured an optical transition frequency at 688 THz in $^{171}$Yb$^+$ with a cesium atomic clock at two times separated by 2.8 years and 
find a value for the fractional variation of the frequency ratio
$f_{\rm Yb}/f_{\rm Cs}$ of
$(-1.2\pm 4.4)\cdot 10^{-15}$~yr$^{-1}$, consistent with zero.
Combined with recently published values 
for the constancy of other transition frequencies
this measurement sets an upper limit on the present variability of $\alpha$ at the level of $2.0\cdot 10^{-15}$~yr$^{-1}~(1\sigma)$, corresponding so far to the most stringent limit from laboratory experiments.
\end{abstract}

\maketitle

Motivation to search for 
a time- or space-dependence of fundamental `constants'  
has developed because theories attempting to unify the fundamental interactions allow or even imply variations of the coupling constants \cite{barrow}.
According to the inflationary model, dramatic changes of particle
properties took place
in the early evolution of the universe and it is conceivable that remnants
of these changes are still observable in later epochs.
Variations of the constants have been searched for in various contexts \cite{uzan,acfc}.
Sommerfeld's fine structure constant $\alpha$
is presently the most important test case. As a dimensionless quantity, $\alpha \simeq 1/137$ can be determined without reference to a specific system of units.
This point is important because the realization of the units themselves may also be affected by variations of the constants.

The only indication for a possible variation of $\alpha$ so far comes from an apparent shift of wavelengths of specific absorption lines produced by interstellar clouds in the light from distant quasars \cite{webb1}. These observations
seem to suggest that about $10^{10}$ yr ago
the value of $\alpha$ was smaller than today: $\Delta \alpha / \alpha= (-0.54\pm 0.12)\cdot 10^{-5}$ 
over the redshift range $0.2<z<3.7$,
representing  $4.7\sigma$ evidence for a varying $\alpha$ 
\cite{webb2}.
Assuming a linear increase in $\alpha$ with time, this would correspond to a drift rate
$\partial \ln \alpha / \partial t=(6.4\pm 1.4)\cdot 10^{-16}~{\rm yr}^{-1}$ \cite{webb2}.
Recent evaluations of new data on quasar absorption lines are consistent with $\Delta \alpha=0$ \cite{sria}. 
A stringent limit on the order of 
$|\partial \ln \alpha / \partial t|<1\cdot 10^{-17}~{\rm yr}^{-1}$
for the temporal variability of $\alpha$ has been deduced from the analysis of a natural nuclear reactor that was active $2\cdot 10^9$ years ago in the Oklo region in West Africa \cite{oklo,uzan}.
Since this analysis is based on the energy of a neutron capture resonance, it is not possible to separate the influence of a variable fine structure constant from possible variations of the coupling constants of the strong interaction and of nuclear masses.
In unified theories the drift rates of these parameters are correlated 
and it has been predicted that nuclear masses and moments may show significantly higher relative drift rates than $\alpha$ \cite{marciano,fritzsch,Karsh2}.
In this sense the limit on 
$\partial \ln \alpha / \partial t$ from the Oklo reactor has to be regarded as model-dependent \cite{cjp}.

In this letter we report a model-independent limit for the present temporal variation of $\alpha$ from comparisons of atomic frequency standards.
We reach a sensitivity for $\Delta \alpha /\Delta t$ with $\Delta t \approx 3$~yr that approaches that of the quasar observations \cite{webb2,sria} if a linear change of $\alpha $ is assumed over a cosmological time span.
Possible systematic errors can usually be identified more easily in laboratory experiments than in astrophysical or geophysical observations.
Precise frequency comparisons have recently been performed in order to
constrain temporal derivatives of fundamental constants
\cite{marion,bize} (see also \cite{uzan} and references therein).
Since these measurements involve 
hyperfine frequencies, they are sensitive not only to a variation of $\alpha$, but also to a variation of the nuclear
magnetic moment.
In this case a limit on the variability of $\alpha$ could only be obtained under the assumption that the nuclear properties stay constant, which is difficult to justify.

To avoid the influence of nuclear structure and to
obtain a direct measure of the drift rate of $\alpha$ one can compare different optical electronic transition frequencies. 
So far, sufficiently accurate atomic optical frequency standards have not been operated continuously over extended periods of time, but a number of
precise frequency measurements have been performed with reference to cesium clocks.
For the analysis, we assume that the electronic transition frequency can be expressed as 
\begin{equation}
f=Ry \cdot C\cdot F(\alpha)
\end{equation}
where 
$Ry=m_e e^4/(8\epsilon_0^2 h^3)\simeq 3.2898\cdot 10^{15}$ Hz 
is the Rydberg constant in SI units of frequency, appearing here as the common atomic scaling factor
($m_e$: electron mass; $e$: elementary charge; $\epsilon_0$: electric constant; $h$: Planck's constant).
$C$ is a numerical constant which depends only on the quantum numbers characterizing the atomic state
and is assumed not to vary in time. $F(\alpha)$ is a dimensionless function 
of $\alpha$
that takes into account level shifts due to relativistic effects \cite{dzu1,dzu2}. 
In  many-electron atoms these effects lead to different sensitivities of the transition frequencies to a change of $\alpha$, with the general trend to find a stronger dependence in heavier atoms, as was first pointed out in 
\cite{prest}. 
The relative temporal derivative of the frequency $f$ can be written as
\begin{equation}
\frac{\partial \ln f}{\partial t} =  \frac{\partial \ln Ry}{\partial t} + 
A \cdot \frac{\partial \ln \alpha}{\partial t}~~~~~~{\rm with}~~~A\equiv \frac{\partial \ln F}{\partial \ln \alpha}.
\end{equation}       
The first term $\partial \ln Ry / \partial t$ represents a variation that would be common to all measurements of electronic transition frequencies: a change of the numerical value of the Rydberg constant in SI units, i.e.
with respect to the $^{133}$Cs hyperfine splitting.
We distinguish here between the numerical value of $Ry$ and the physical parameter which determines the atomic energy scale. A change of the latter would not be detectable in this type of measurement because the   
$^{133}$Cs hyperfine splitting is also proportional to this quantity.
Consequently, the detection of a nonzero value of $\partial \ln Ry / \partial t$ would have to be interpreted as being due to a change of the cesium hyperfine interval, and  in particular of the $^{133}$Cs nuclear magnetic moment.
The second term in Eq. 2 is specific to the atomic transition under study.
The value of the sensitivity factor $A$ for small changes of $\alpha$ has been calculated for a number of cases
by  Dzuba et al. \cite{dzu1,dzu2}.         
Assuming that possible drifts of the atomic frequencies are linear in time over the period of interest, the drift rate $\partial \ln \alpha / \partial t$ can be obtained from at least two measured drift rates of transition frequencies $\partial \ln f / \partial t$ if the sensitivity factors $A$ for the two transitions are different. 
A plot of $\partial \ln f / \partial t$ as a function of $A$ should yield a straight line where the slope is determined by  $\partial \ln \alpha / \partial t$ and a drift of $Ry$
would show up in the intercept \cite{Karsh2}.

\begin{figure}[tbh]
\begin{center}
\includegraphics[width=8cm]{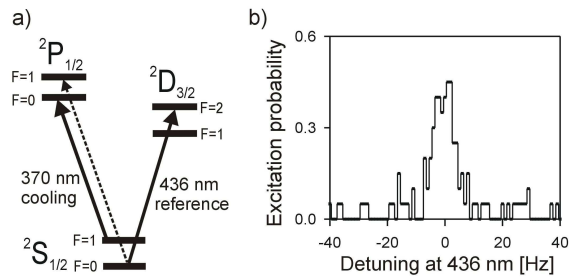}
\end{center}
\caption{a) Simplified level scheme of the $^{171}$Yb$^+$ ion. 
A modulation sideband of the cooling laser is used to repump the ion from the $F=0$ ground state.
b) Excitation spectrum of the $^2S_{1/2}(F=0,m_F=0)\rightarrow {^2D}_{3/2}(F=2,m_F=0)$ transition,
obtained with probe pulses of 90 ms duration 
and 20 probe cycles for each value of the detuning.}
\end{figure}

Optical transition frequencies can be measured very precisely in single trapped and laser-cooled ions \cite{Madej}
because here perturbations that may lead to systematic frequency shifts can be controlled very effectively. These experiments are well suited for a laboratory search for a variation of $\alpha$ because 
transitions in heavy ions like Yb$^+$ or Hg$^+$ cover a wide range of $A$ values \cite{dzu1,dzu2}.
At PTB an optical frequency standard based on  $^{171}$Yb$^+$ is investigated \cite{Tamm1,icols}.
A single ytterbium ion is trapped in a miniature Paul trap and is laser-cooled to a sub-millikelvin temperature
by exciting the $^2S_{1/2}(F=1) \rightarrow {^2P_{1/2}}(F=0)$ resonance transition at 370 nm.
Two additional lasers at wavelengths of 935~nm and 639~nm are used to repump the ion from the metastable
$^2D_{3/2}(F=1)$ and $^2F_{7/2}$ levels that may be populated during the cooling phase.
The electric quadrupole transition $6s\,^2S_{1/2}(F=0) \rightarrow 5d\,^2D_{3/2}(F=2)$
at 436 nm wavelength (688 THz) with a natural linewidth of 3.1~Hz serves as the reference transition. 
The isotope $^{171}$Yb has a nuclear spin quantum number of $1/2$ so that a $(F=0,m_F=0)\rightarrow (F=2,m_F=0)$ component of the reference transition with vanishing linear Zeeman shift is available (Fig. 1a).
The reference transition is probed by the frequency-doubled radiation from a diode laser
emitting at 871 nm. The short-term frequency stability of this laser is derived from a high-finesse, temperature-stabilized and seismically isolated reference cavity.
The trapped ion is interrogated using the electron shelving scheme \cite{Madej} with
alternating cooling and probe laser pulses.
A Fourier-limited linewidth of 10 Hz at 688 THz 
and a resonant excitation probability of about 50\% is obtained (Fig. 1b). 
The probe laser frequency is stabilized to the atomic reference frequency in a servo system with a time constant of $\approx$ 10 s. 
We have observed a fractional instability well below $1\cdot 10^{-15}$ for an averaging time of 1000 s.
In order to minimize servo errors due to the drift of the probe laser frequency (in the range of 0.2 Hz/s), a second-order integrating algorithm is used.

The frequency of the 688 THz $^{171}$Yb$^+$ standard was  measured relative to PTB's primary cesium fountain clock \cite{BauchCs}.
A hydrogen maser is referenced to the cesium clock and delivers a signal at 100 MHz with an instability below $2\cdot 10^{-15}$ in 1000 s of averaging.
The link between the maser signal and 
the optical frequency is established 
with a femtosecond-laser frequency comb generator \cite{udem}. 
The result of a first series of measurements starting on MJD 51889
was $688\, 358\, 979\, 309\, 312$~Hz with a total $1\sigma$ measurement uncertainty of 
$6$ Hz \cite{Stenger1}.
The frequency was remeasured around MJD 52947 after improvements in the resolution of the ionic resonance and 
in the frequency comb generator setup. The new value has a total $1\sigma $ uncertainty of 6.2~Hz (see below) and is insignificantly lower by 2.3~Hz (see Fig. 2). From these measurements we deduce
a value for the fractional variation of $f_{\rm Yb}/f_{\rm Cs}$ of
$(-1.2\pm 4.4)\cdot 10^{-15}$~yr$^{-1}$.

\begin{figure}[tbh]
\begin{center}
\includegraphics[width=5cm]{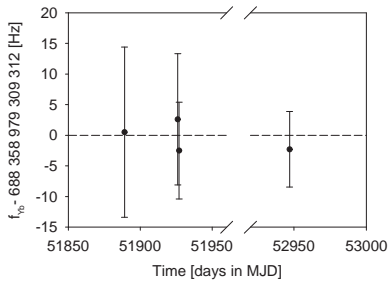}
\end{center}
\caption{Results of absolute frequency measurements of the $^{171}$Yb$^+$ standard versus the date of the measurement
(MJD: Modified Julian Date).}
\end{figure}

The dominant source of systematic uncertainty 
in the $^{171}$Yb$^+$ 688 THz optical frequency standard
is the so-called quadrupole shift, i.e., the shift of the atomic transition frequency due to the interaction
of the electric quadrupole moment of the $D_{3/2}$ state with the gradient of the static electric field at 
the trap center.
Such a field gradient may be produced by patch charges on the trap electrodes
and 
a sensitivity of the transition frequency to field gradients of up to
6~Hz/(V/mm$^2$) is expected \cite{icols}.
A static field gradient also leads to modifications of the oscillation frequencies of the ion in the trap
and an analysis of this effect indicates that patch field gradients in our trap are smaller 
than 0.5~V/mm$^2$.
Other systematic frequency shifts are well below 1 Hz and their contribution to the uncertainty is negligible at present: The static magnetic field in the trap is typically 1 $\mu$T during the
interrogation of the reference transition, leading to a quadratic Zeeman shift of 0.06 Hz.  
The quadratic Stark shift due to blackbody radiation emitted from the trap electrodes at room
temperature is calculated as $-0.4$~Hz.
In the absence of a full evaluation,
we use an estimate of 3.2~Hz for the systematic uncertainty
of the $^{171}$Yb$^+$ frequency standard, dominated by the contribution from the quadrupole shift.
Frequency comparisons between two Yb$^+$ ions in two separate traps showed agreement at the sub-hertz level \cite{icols}.
The total uncertainty of 6.2~Hz (corresponding to a relative uncertainty of $9\cdot 10^{-15}$) contains also
a systematic contribution of 2.2~Hz from the microwave reference and frequency comb generator and a total statistical contribution of 4.8~Hz.

To obtain the limit on $\partial \ln \alpha / \partial t$ we combine the data from $^{171}$Yb$^+$ with a published 
limit on the drift rate of a transition frequency in $^{199}$Hg$^+$ \cite{bize}.
The mercury ion is investigated as an optical frequency standard at NIST (Boulder, USA).
The  transition $5d^{10}6s~^2S_{1/2}\rightarrow 5d^96s^2~^2D_{5/2}$ at 282~nm (1065 THz) serves as the reference. 
A first frequency measurement of this transition was published in 2001 with a relative uncertainty of $9\cdot 10^{-15}$ \cite{Udem2}.
A sequence of measurements over a period of two years has resulted in a constraint on the fractional variation of  
$f_{\rm Hg}/f_{\rm Cs}$ at the level of $7\cdot 10^{-15}$~yr$^{-1}~(1\sigma)$ \cite{bize}. 

The transition frequencies of the ytterbium and of the mercury ion are the two most accurately known optical frequencies
and their sensitivities to changes of  $\alpha$ are quite different \cite{foot}: $A_{\rm Yb}=0.88$  \cite{dzu2} 
and $A_{\rm Hg}=-3.19$ \cite{dzu1}. 
Since the uncertainties of these numbers are not given explicitly, we treat them as being exact, noting that changes within 10\% would modify our results only marginally.
Combining the two measurements in a search for a temporal variability of $\alpha$
is well justified:
The Hg$^+$ observations cover the time between MJD 51770 and MJD 52580 so that there is a significant temporal overlap. 
The cesium fountains of NIST and of PTB have been compared repeatedly during the relevant period and found to be in agreement to within their evaluated uncertainties of $1.0\cdot 10^{-15}$ \cite{BauchCs}.    

A further precision test for the constancy of an optical transition frequency was recently presented by
Fischer et al. who measured the $1S\rightarrow 2S$ transition frequency in atomic hydrogen in two periods around MJD 51360 and MJD 52680, using a transportable cesium fountain frequency standard \cite{fischer}.
The limit on the fractional drift rate for the hydrogen frequency that is deduced from these experiments is $(-3.2\pm 6.3)\cdot 10^{-15}$~yr$^{-1}$ \cite{fischer} 
with $A_{\rm H}$ being zero.

The three limits on the frequency drift rates are consistent with a linear dependence on $A$, as assumed in Eq. 2 (see Fig. 3a).
Through a weighted linear regression we obtain from the slope of the fit:
\begin{equation}
\frac{\partial \ln \alpha}{\partial t} = (- 0.3 \pm 2.0)\cdot 10^{-15}~{\rm yr}^{-1}
\end{equation}
for the present value of the temporal derivative of the fine structure constant at the confidence level of $1\sigma$. 
Using only the data points from the single-ion experiments with
Hg$^+$ and Yb$^+$ one finds only a marginal change of the result:
$\partial \ln \alpha /\partial t = (- 0.2 \pm 2.0)\cdot 10^{-15}$~yr$^{-1}$.
This limit is the result of 
a model-independent analysis that is not based on assumptions about correlations between drift rates of different quantities and does not postulate the drift rate of any quantity to be zero. To our knowledge, this is so far the most stringent limit on $\partial \ln \alpha/ \partial t$ obtained from laboratory experiments.
Another recent related analysis based on the data from hydrogen and mercury is presented in \cite{fischer}.

\begin{figure}[tbh]
\begin{center}
\includegraphics[width=5cm]{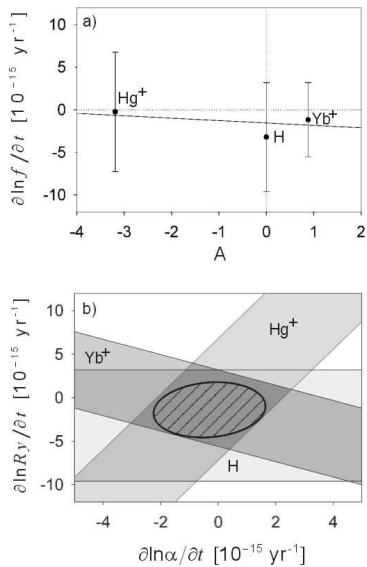}
\end{center}
\caption{a) Relative frequency drift rate versus sensitivity factor $A$ for the $S\rightarrow D$ reference transitions in the Hg$^+$ and Yb$^+$ ions and for the $1S\rightarrow 2S$ transition in atomic hydrogen. The solid line is the result of a weighted linear regression.
A significant deviation of the slope from zero would indicate a nonzero 
time derivative of the  fine structure constant $\alpha$.
b) Constraints on the variability of $\alpha$ and $Ry$ as obtained from the experimental points in (a). 
Stripes mark the 1$\sigma$ uncertainty of the individual measurements
and the central hatched area is bounded by the standard uncertainty ellipse resulting from the 
combination of all three experiments.}
\end{figure}

The intercept of the fitted straight line with the line $A=0$ in Fig. 3a  can be used to determine the drift rate of the numerical value of the Rydberg frequency $Ry$.
This quantity is of great metrological importance.
A nonzero value would imply among other consequences that a time scale generated with a cesium clock
would not be uniform with respect to other atomic oscillations.
The result that we obtain here is consistent with zero:  
\begin{equation}
\frac{\partial \ln Ry}{\partial t}=(-1.6 \pm 3.2)\cdot 10^{-15}~{\rm yr}^{-1}.
\end{equation}
The result for  the fit based only on the single-ion data is $\partial \ln Ry/\partial t=(-1.0 \pm 3.7)\cdot 10^{-15}$~yr$^{-1}$. 

Fig. 3b) shows how the measurements of the three transition frequencies contribute to the constraints on the temporal derivatives of $\alpha$ and $Ry$. The central hatched area is bounded by the standard uncertainty ellipse as defined by 
\begin{equation}
\sum_{i} \frac{1}{u_i^2}\left(  
\frac{\partial \ln f_i}{\partial t}-\frac{\partial \ln Ry}{\partial t}-A_i\frac{\partial \ln \alpha}{\partial t}\right) ^2 =1+\chi^2_{\rm min}
\end{equation}
where $i$ designates the elements (H, Yb, Hg), $\partial \ln f_i/\partial t$ is the central value of the observed drift rate, $u_i$ its $1\sigma$ uncertainty
and $\chi^2_{\rm min}=0.092$ the minimized $\chi^2$ of the fit.

The kind of analysis that we have applied here may be extended to other atomic systems as soon as precision data on frequency drift rates become available. 
Further improvements can be expected since trapped ions have the potential to reach 
relative uncertainties in the range of $10^{-18}$ for the comparison of optical transition frequencies \cite{Madej}.
The combination of results from different atomic and molecular frequency standards will make it possible to disentangle the contributions from the fundamental interactions and thus lead to sensitive tests of the consequences of unification theories.

We thank S. Weyers, R. Wynands and A. Bauch for providing the
Cs-fountain and H-maser 
reference and for helpful discussions. 
The work was supported by DFG through SFB 407 and the work of S.G.K. in part by RFBR
under grants 03-02-16843.


\begin{thebibliography}{99}

\bibitem{barrow} J. D. Barrow, Astrophys. Space Sci. {\bf 283}, 645 (2003).     

\bibitem{uzan} J.-P. Uzan, Rev. Mod. Phys. {\bf 75}, 403 (2003).

\bibitem{acfc} {\it Astrophysics, Clocks and Fundamental Constants}, Eds.: S. G. Karshenboim and E. Peik,
(Springer, Heidelberg, 2004).

\bibitem{webb1} J. K. Webb et al., Phys. Rev. Lett. {\bf 87}, 091301 (2001).

\bibitem{webb2} M. T. Murphy, J. K. Webb, V. V. Flambaum, Mon. Not. R. Astron. Soc. {\bf 345}, 609 (2003).

\bibitem{sria}R. Quast, D. Reimers, S. A. Levshakov, Astr. Astrophys. {\bf 415}, L7 (2004); R. Srianand et al., Phys. Rev. Lett. {\bf 92}, 121302 (2004).

\bibitem{oklo} A. I. Shlyakhter, Nature {\bf 264}, 340 (1976);
T. Damour, F. Dyson, Nucl. Phys. B {\bf 480}, 596 (1994).

\bibitem{marciano}W. J. Marciano, Phys. Rev. Lett. {\bf 52}, 489 (1984).

\bibitem{fritzsch}X. Calmet, H.  Fritzsch, Eur. Phys. J. C {\bf 24}, 639 (2002);
P. Langacker, G. Segre, M. J. Strassler, Phys. Lett. B {\bf 528}, 121 (2002).

\bibitem{Karsh2} S. G. Karshenboim, eprint physics/0311080.

\bibitem{cjp} S. G. Karshenboim, Can. J. Phys. {\bf 78}, 639 (2001).

\bibitem{marion} H. Marion et al., Phys. Rev. Lett. {\bf 90}, 150801 (2003).

\bibitem{bize} S. Bize et al., Phys. Rev. Lett. {\bf 90}, 150802 (2003).

\bibitem{dzu1} V. A. Dzuba, V. V. Flambaum, J. K. Webb, Phys. Rev. A {\bf 59}, 230 (1999).

\bibitem{dzu2} V. A. Dzuba, V. V. Flambaum, M. V. Marchenko, Phys. Rev. A {\bf 68}, 022506 (2003).

\bibitem{prest} J. D. Prestage, R. L. Tjoelker, L. Maleki, Phys. Rev. Lett. {\bf 74}, 3511 (1995).

\bibitem{Madej}
A. A.~Madej, J. E.~Bernard, in {\it Frequency Measurement and Control:
Advanced Techniques and Future Trends}, Springer Topics in Applied
Research, Ed.: A. N.~Luiten (Springer, Berlin, Heidelberg, 2000).

\bibitem{Tamm1}
Chr. Tamm, D. Engelke, and V. B\"uhner, Phys. Rev. A {\bf 61}, 053405 (2000).

\bibitem{icols} Chr. Tamm, T. Schneider, and E. Peik, in
{\it Laser Spectroscopy XVI}, Eds.: P. Hannaford, A.
Sidorov, H. Bachor, and K. Baldwin (World Scientific, Singapore, 2004).

\bibitem{BauchCs} A. Bauch, Meas. Sci. Technol. {\bf 14}, 1159 (2003); T. E. Parker et al., Proc. 2001 IEEE Intl. Freq. Contr. Symp., p. 63.

\bibitem{udem} Th. Udem, R. Holzwarth, T. W. H\"ansch, Nature {\bf 416}, 233 (2002).

\bibitem{Stenger1} J. Stenger et al., Opt. Lett. {\bf 26} 1589 (2001).

\bibitem{Udem2} Th. Udem et al.,  Phys. Rev. Lett. {\bf 86}, 4996 (2001).

\bibitem{foot}
The change of sign between the two elements reflects the fact that in Yb$^+$ a 6s-electron is excited to the empty 5d-shell, while in Hg$^+$ a hole is created in the filled 5d-shell if the electron is excited to 6s.

\bibitem{fischer} M. Fischer et al., Phys. Rev. Lett. {\bf 92}, 230802 (2004).
 
\end{thebibliography}
\end{document}